\documentclass[useAMS,usegraphicx]{mn2e}

\usepackage{aas_macros}
\usepackage{bm}
\usepackage{commath}
\def\gs{\mathrel{\raise0.35ex\hbox{$\scriptstyle >$}\kern-0.6em
\lower0.40ex\hbox{{$\scriptstyle \sim$}}}}
\def\ls{\mathrel{\raise0.35ex\hbox{$\scriptstyle <$}\kern-0.6em
\lower0.40ex\hbox{{$\scriptstyle \sim$}}}}
\renewcommand{\d}[1]{\ensuremath{\operatorname{d}\!{#1}}}

\title[Radio luminosity functions from stack-fitting]
{Cosmic star formation probed via parametric stack-fitting of known sources to radio imaging}
\author[I.G.~Roseboom]
{\parbox{\textwidth}{\raggedright I.G.~Roseboom$^{1}$\thanks{E-mail: \texttt{igr@roe.ac.uk}}
 and P.N.~Best$^{1}$}\vspace{0.4cm}\\
\parbox{\textwidth}{\raggedright
$^{1}$Institute for Astronomy, University of Edinburgh, Royal Observatory, Blackford Hill, Edinburgh EH9 3HJ, UK\\
}}

\begin{document}

\date{\today}

\pagerange{\pageref{firstpage}--\pageref{lastpage}} \pubyear{2011}

\maketitle

\label{firstpage}

\begin{abstract}
The promise of multi-wavelength astronomy has been tempered by the large disparity in sensitivity and resolution between different wavelength regimes. Here we present a statistical approach which attempts to overcome this by fitting parametric models directly to image data. Specifically, we fit a model for the radio luminosity function (LF) of star-forming galaxies to pixel intensity distributions at 1.4\,GHz coincident with near-IR selected sources in COSMOS. Taking a mass-limited sample in redshift bins across the range $0<z<4$ we are able to fit the radio LF with $\sim0.2$\,dex precision in the key parameters (e.g. $\Phi^{\ast}$,$L^{\ast}$).  Good agreement is seen between our results and those using standard methods at radio and other wavelengths. Integrating our luminosity functions to get the star formation rate density we find that galaxies with $M_{\ast}>10^{9.5}$\,M$_{\odot}$ contribute $\gs50$ per cent of cosmic star formation at $0<z<4$. The scalability of our approach is empirically estimated, with the precision in LF parameter estimates found to scale with the number of sources in the stack, $N_{\rm s}$, as $\propto \sqrt{N_{\rm s}}$. This type of approach will be invaluable in the multi-wavelength analysis of upcoming surveys with the SKA pathfinder facilities; LOFAR, ASKAP and MeerKAT.

\end{abstract}


\section{Introduction}
The cosmic history of star formation, typically summarised as the redshift evolution of the star formation rate density (SFRd; Madau et al.\ 1996; Lilly et al.\ 1996; Hopkins \& Beacom 2006; Behroozi et al. 2013), is widely regarded as one of the benchmark measurements of galaxy evolution; its precise measure is a key goal of modern observational astrophysics, and any model for galaxy formation must replicate these observations to be taken seriously.

The rapid evolution of observational facilities in the last decade, in particular upgrades to the wide-field optical (with the Advanced Camera for Surveys in 2002) and near-IR (with Wide Field Camera 3 in 2009) capabilities of the {\it Hubble Space Telescope} (HST), has resulted in robust estimates of the SFRd out to at least $z=7$ (Bouwens et al. 2006, 2008, 2011; McClure et al.\ 2010, 2013). However, the HST view of the high-$z$ Universe is effectively limited to rest-frame UV and optical wavelengths and hence to relatively unobscured galaxies. This limitation is worrisome, as populations of highly obscured star-forming (100--1000\,M$_{\odot}$\,yr$^{-1}$) galaxies are seen in the far-IR and submm (e.g. Hughes et al.\ 1999). Large scale submm surveys have shown that the number density of these so-called Ultra-luminous IR galaxies (ULIRGs) are dwarfed by the more moderate star-forming population at all redshifts (e.g. Reddy et al.\ 2008), although they do occur in large enough quantities to cause tension with numerical models of galaxy formation (e.g. Baugh et al.\ 2006; Lacey et al.\ 2010). Existing far-IR/submm facilities do not possess the sensitivity to probe even moderate SFRs (i.e. SFR$<100$\,M$_{\odot}\,$yr$^{-1}$) at $z>1$ and hence the overlap in UV/optical and far-IR/submm identified galaxies at high redshift is small. 

This situation is clearly unsatisfactory; to observe the full range in SFRs at high redshift requires measurements at two distinct wavelengths which can only be cross-calibrated in the nearby Universe (i.e. $z<0.2$). Deep, wide-area observations in the UV/optical would offer the observationally cheapest route to reconciling this disconnect. However, the strong relationship between SFR and dust obscuration means that the galaxies with the highest star formation rates may be amongst the faintest at the UV/optical wavelengths (e.g. Meurer et al.\ 1999; Hopkins et al. 2001; Kewley et al.\ 2002; Reddy et al.\ 2010; Roseboom et al.\ 2012). Moreover, there is some evidence that UV/optical tracers of the level of dust obscuration (e.g. the UV slope, $\beta_{\rm UV}$, the ratio of Balmer lines) may only be accurate at moderate SFRs (i.e. $<100$\,M$_{\odot}\,$yr$^{-1}$; Reddy et al.\ 2010; Buat et al.\ 2010; Roseboom et al.\ 2012). 

Thus there is no alternative but to push the sensitivity limits of SFR tracers which are not affected by dust obscuration. In recent years it has become de rigueur to build vast multi-wavelength datasets on a number of well-known extragalactic survey fields (e.g. GOODS, Lockman Hole, COSMOS, etc). While the disparity in observational capabilities across the electromagnetic spectrum has limited the usefulness of multi-wavelength studies of individual objects, the practice of stacking less sensitive datasets (typically X-ray, far-IR/submm, radio) at the positions of the deepest catalogued sources (typically optical/near-IR), has become common-place. For imaging afflicted by white noise the stacked signal-to-noise should increase proportional to the square-root of the number of sources in the stack. Hence for extra galactic fields with 10,000s of sources the aggregate, dust unbiased, SFRs can be robustly estimated at a level $\sim100$ times fainter than the detection limit for individual sources. 

However, traditional stacking methods can only recover the mean (or median) SFR of any sample, and hence only the mean relationship between SFR and the binned variable can be established. To counter this it is common to bin the sources in more than a single parameter, e.g. the stellar mass - redshift plane (Oliver et al.\ 2010, Karim et al.\ 2011, Roseboom et al.\ 2013). This approach has the significant drawback of reducing the number of sources, $N$, in each stack, which in turn reduces the stacked signal by a factor proportional to $N^2$.

In this paper we present a statistical framework that allows the underlying distributions i.e. $P({\rm SFR}|M_{\ast}, z, \beta_{\rm UV}, \ldots)$ to be extracted in a parametric way from the stack. The clear advantage of this approach is that it allows the distribution in SFR to be constrained using all the available sources in a consistent way. To illustrate this point, Karim et al.\ (2011; K11) perform a stacking analysis with practically the same dataset we will use here;  optical/near-IR selected galaxies stacked into VLA imaging in COSMOS. They split their sample into $9\times7$ redshift-mass bins and determine the mean SFR in each bin via median stacking of cutouts of the radio data around each source. If no binning is required, and the precision in the parametric fit scales as $\sqrt{N}$, then our method could provide estimates of the SFR-$M_{\ast}$-$z$ relation with $\sim 8$ times more precision ($\sim 3$ if we still bin in redshift). These types of gains are non-negligible considering the observational expensive required to improve the raw sensitivity of large multi-wavelength datasets like COSMOS by these factors.


In \S\ref{sec:stackmethod} is presented the method used to extract $P({\rm SFR}|z)$ from a stack of galaxy positions, \S\ref{sec:data} describes the datasets used to test this method, and \S\ref{sec:sresults} presents our results. \S\ref{sec:sfrd} compares our results to the literature and discusses the implications of our results in the context of the cosmic star formation rate density, while \S\ref{sec:future} discusses the potential future applications for this method, and possible obstacles. Finally, \S\ref{sec:conc} presents our conclusions.

Throughout we assume a $\Lambda$CDM cosmology with
$\Omega_{\Lambda}=0.7$, $\Omega_{\rm m}=0.3$ and
$H_0=70$\,km\,s$^{-1}$\,Mpc$^{-1}$. Where relevant all quoted quantities assume a Chabrier (2003) IMF.

\section{Parametric stacking method}\label{sec:stackmethod}
\subsection{Basic framework}
If we consider a pixelated sky image ${\bm d}$ in units of flux density per telescope beam then for a list of sources with known positions ${\bm x}$ the best estimate of the mean flux density will be given by 
\[\frac{\sum {\bm d_x}/{\bm \sigma_x^2}}{\sum 1/{\bm \sigma_x^2}}, \]

\noindent where ${\bm d_x}$ and ${\bm \sigma_x}$ are the pixel flux density and noise estimates for positions ${\bm x}$, respectively. Turning this around, if the true flux densities at positions ${\bm x}$ are ${\bm f_x}$  the observed pixel intensities $d_x$ will be simply,
\[{\bm d_x}={\bm f_x} + \delta,\]
where $\delta$ is some noise value drawn from a Gaussian distribution; $N(\mu=0,{\bm \sigma_x})$. If a model $M$ can be constructed which predicts the probability $P(f)$ of a true flux density $f$ for a known source then the probability of observing a particular pixel intensity $d$ at the location of that source is,

\begin{equation}
P(d | M,\sigma) = \int_{0}^{\infty} \frac{1}{\sigma\sqrt{2\pi}}P(f)\exp\left(\frac{-(f-d)^2}{2\sigma^2}\right)\d{f},
\label{Eqn:pdm}
\end{equation}

\noindent and so for a stack of sources ${\bm x}$,
\begin{equation}
P({\bm d_x} | {\bm x}, M, {\bm \sigma_x}) = \Pi_{\rm x} P(d | M,\sigma).
\label{Eqn:pdxm}
\end{equation}

\noindent Applying Bayes' theorem we see that,
\[P(M, {\bm \sigma_x |{\bm d_x}) \propto P(M)P(\bm d_x} | {\bm x}, M, {\bm\sigma_x}),\]
\noindent and so the best-fit parameters for the model $M$ can be found by maximising the likelihood given in Eqn. \ref{Eqn:pdxm}.

\subsection{Building the flux density distribution model}
Equations \ref{Eqn:pdm} and \ref{Eqn:pdxm} give a generic framework to fit a stack (i.e. stack-fit) of flux densities for a set of known positions in the presence of noise. This framework could be used for parameter estimation in models with a wide range of purposes, from simple parameterisations of the number counts (e.g. Mitchell-Wynne et al.\ 2013) to full-blown galaxy formation simulations (e.g. Henriques et al.\ 2010). Here we propose to model the rest-frame luminosity distribution function at a given redshift, $z$; $\hat{\Phi}(l,z)$. In this scenario we need to convert our proposed luminosities to the observed flux density in order to use Eqn.~\ref{Eqn:pdxm}. Considering both the luminosity distance and a k-correction as a function of redshift then,
\begin{equation}
P(f) = \frac{\hat{\Phi}(4\pi D(z)^2k(z)f,z)}{\int_{l=0}^{\infty}\hat{\Phi}(l,z)\d{l}\d{z}} 
\label{Eqn:pfz}
\end{equation}

\noindent where $D(z)$ is the luminosity distance and $k(z)$ the k-correction at a redshift $z$. It is worth emphasising that $P(f)$ here is the probability of a {\it particular} galaxy at a known position having a true flux density $f$. If for a given sample $\hat{\Phi}$ is independent of redshift then we can simply propose $\hat{\Phi}(l)$ and remove the integral over $\d{z}$ in the denominator of Eqn.~\ref{Eqn:pfz}. Similarly, if we can split our sample into narrow redshift bins then the simplest non-parametric estimate of the redshift evolution can be obtained by measuring $\hat{\Phi}(l)$ in each redshift bin independently. This is the approach we will take later in \S\ref{sec:phimod}.

Equation \ref{Eqn:pfz} is analogous to the formula for $P(M_{\mathrm abs},z)$, where $M_{\mathrm abs}$ is the absolute magnitude, which appears in the Sandage, Tammann \& Yahil (1979; STY) maximum likelihood LF estimator. However in our scenario the sample is not selected at the wavelength at which $\hat{\Phi}$ is determined, but by a selection at some other wavelength (or physical galaxy property). In the STY formalism the integral in the denominator of Eqn.~\ref{Eqn:pfz} would have a lower bound equal to the minimum observable luminosity at the redshift of the source. However, in our formalism the sources in the stack can have no such limit and so the lower bound of this integral is zero. This means that whatever parametric form we choose for $\hat{\Phi}$ must have a finite integral from zero to infinity, otherwise the ratio in Eqn.~\ref{Eqn:pfz} will become undefined.

If the redshifts for each source are not known accurately, or only the redshift distribution for the population of sources is known (as is the case for e.g. BzK galaxies), then the probability density of the redshift, $P(z)$, can be integrated over, i.e. 
\begin{equation}
P(f) = \frac{1}{\int_{l=0}^{\infty}\hat{\Phi}(l,z)\d{l}\d{z}} \int_{z=0}^{\infty} P(z)\hat{\Phi}(4\pi D(z)^2k(z)f,z)\d{z}.
\label{Eqn:pfdz}
\end{equation}

Similarly, if the k-correction cannot be described as a simple function of redshift then whatever other parameters, ${\bm \eta}$, that are required to describe the k-correction must be added to the integral, i.e. $\int\int k(z,{\bm \eta})\d{z}\d{\bm \eta}$.

As is the case with the STY LF estimator, Eqn.~\ref{Eqn:pfz} is insensitive to the normalisation of the LF model $\hat{\Phi}$. The value of the normalisation can be recovered by considering the total number of observed sources, their visibility, and various incompletenesses, as we will detail in the following \S\ref{sec:norm}.

\subsection{Dealing with incompleteness}
Armed with Eqns. \ref{Eqn:pdm} and \ref{Eqn:pfdz} we have the statistical framework needed to fit a model specified in terms of rest-frame luminosities directly to the pixel stack. It is worth noting that this setup assumes implicitly that the input catalogue to the stack is complete. This is not an onerous requirement, the depth of optical/near-IR surveys is such that it is possible to construct complete volume or mass-limited samples out to high redshift. However, it would be preferable to generalise our method to accommodate known incompleteness in the input catalogues. To accomplish this we need to calculate $P(f|\bm{\theta})$, where $\bm{\theta}$ are the properties of a source in the catalogue providing the positions, e.g. apparent magnitude, stellar mass, photometric redshift, optical colour, etc. If $P(f)$ and $P(\bm{\theta})$ are independent then,

\begin{equation}
P(f|\bm{\theta})=P(f)P(\bm \theta),
\end{equation}

\noindent i.e. we simply multiply $P(f)$ by the estimated completeness for sources with properties $\bm{\theta}$.

\subsection{Determining the normalisation}\label{sec:norm}

In converting our model $\hat\Phi(l)$ to a probability distribution the absolute normalisation, $\hat{\Phi}^{\ast}$, is lost. To recover this we need to determine which value of $\Phi^{\ast}$ will give the correct number of observed galaxies, $N$, i.e.
\begin{equation}
\Phi^{\ast}=\frac{\bar{n}}{\int_{l=0}^{\infty}\hat\Phi(l,z)\d{l}\d{z}},
\end{equation}

\noindent where $\bar{n}$ is the mean density for objects in the stack. For a complete, volume-limited, sample of sources it is clear that,
\[
\bar{n}=\frac{N}{V},
\]
\noindent where $V$ is the observable volume for sources in the stack list. For non-complete stack lists with variable completeness, $C$, and observable volumes, $V_{\rm max}$, this becomes,
\begin{equation}
\bar{n}=\sum_x [C_xV_{\rm max}]^{-1}
\label{Eqn:cvcor}
\end{equation}
\section{Data and Model Setup}\label{sec:data}

\subsection{ULTRAVISTA and VLA data in COSMOS}
To use our method effectively we need two overlapping datasets; an input catalogue which is effectively deeper and provides redshift estimates, and a image which has pseudo-white noise and similar angular resolution to the input catalogue (to avoid issues with confusion). The ULTRAVISTA photometric redshift catalogue of Muzzin et al.\ (2013a) and the VLA-COSMOS 1.4\,GHz radio image presented by Schinnerer et al.\ (2007) satisfy these two requirements, respectively. The Muzzin et al.\ (2013a) catalogue provides photometric redshifts and stellar mass estimates for over 260,000 galaxies to a limiting magnitude of $K_s<24.35$ ($3\sigma$) from the combination of the 1.63 deg.$^2$ ULTRA-VISTA $Y, J, H, K_s$ imaging (McCracken et al. 2012) with overlapping ground-based and HST optical imaging of the COSMOS field (see Scoville et al.\ 2008).  Only reliable sources from the ULTRAVISTA catalogue are considered here by requiring the metrics {\sc star}$=0$, {\sc K\_flag}$<4$, {\sc contamination}$!=1$ and {\sc nan\_contam}$<3$, and a magnitude of $K_s<23.4$, as recommended by Muzzin et al.\ (2013a). These flags allow the removal of stars as well as sources with corrupted photometry due to bad pixels and nearby bright sources. This magnitude limit is chosen as it represents the 90 per cent completeness level of the catalogue.

Meanwhile, the VLA imaging of the COSMOS field consists of a mosaic of 23 tiled pointings of the VLA at 1.4\,GHz in the A and C-configuration (Schinnerer et al.\ 2007). The image has an RMS of  $\sim10\,\mu$Jy\,beam$^{-1}$ at the field centre and a beam FWHM of $1.4\times1.5$\,arcsec. In this work we only make use of the 1.6\,deg.$^2$ of the mosaic that overlap with ULTRAVISTA coverage. The publically available image has been ``cleaned", i.e. point sources above 40$\,\mu$Jy responding to the synthesised (``dirty'') beam have been removed and replaced with a Gaussian beam conserving the peak flux density. For our peak pixel stacks this has no effect, as the peak flux density is conserved in the cleaning process; however this does have some impact on the integrated-to-peak flux density ratio, as we will discuss in \S\ref{sec:zbins}.

\subsection{Modelling the mass-limited radio luminosity function}\label{sec:phimod}
To use Eqn.~\ref{Eqn:pfz} we need to define some parametric form for $\hat{\Phi(l,z)}$. As we have high quality photometric redshifts for all sources in the input catalogue we prefer to divide our sample into redshift bins across which the evolution of the LF is negligible and fit for $\hat{\Phi(l)}$ in each of these redshift bins independently. This allows the redshift dependance of the LF to be probed in a non-parametric way. Typically the radio luminosity function of star-forming galaxies is assumed to follow a modified Schecter function (Saunders et al.\ 1990);
\begin{equation}
\hat{\Phi}(l)={\Phi}^{\ast}\left(\frac{l}{l^{\ast}}\right)^{1-\alpha}\exp\left[-\frac{1}{2\sigma_{\mathrm LF}^2}\log^2_{10}\left(1+\frac{l}{l^{\ast}}\right)\right]\frac{\d{l}}{l}.
\label{Eqn:msf}
\end{equation}

Motivated by this, and to allow direct comparison with literature estimates, we also assume that the luminosity distribution for our mass-limited sample in each redshift bin is described by Eqn.~\ref{Eqn:msf}. As the radio emission in these star-forming galaxies will be dominated by the synchrotron radiation we assume the radio SED can be described as  $S_{\nu}\propto\nu^{-\alpha_s}$. Thus the k-correction is given by $k(z)=(1+z)^{1-\alpha_s}$. Here we assume $\alpha=0.8$ (Condon et al.\ 1992).

Finally, as we are will be working with a mass-limited sample we require that the integral of $\hat{\Phi}(l,z)$ be finite, i.e. $\alpha<1$. 

\noindent  

\subsection{Redshift binning and model fitting}\label{sec:zbins}
We build samples for input to our stack-fitting method by slicing the ULTRAVISTA catalogue into nine redshift bins across the range $0.1<z<4$. In order to restrict the sample to only star forming galaxies, i.e. excluding AGN and passive galaxies, we use the $UVJ$ colour selection first described by Williams et al.\ (2009). Specifically we require rest-frame $[U-V]\le0.88[V-J]+c$ where $c=0.69$ at $z<1$, and $c=0.59$ at $z>1$. Table \ref{tab:stackbins} details the redshifts and numbers of galaxies in our redshift bins. As an additional restriction we consider only sources with stellar mass $M_{\ast}>10^{9.5}\,$M$_{\odot}$, the $\gs50$ percent completeness limit for the ULTRAVISTA sample at the highest redshifts we consider.

For each source in the ULTRAVISTA catalogue we calculate a completeness to use in Eqn.~\ref{Eqn:cvcor}. The completeness correction is estimated by comparing the number of observed sources in a mass bin (with $\Delta M_{\ast}=0.5$) to the predicted number taking into account the geometry of the ULTRAVISTA survey, the redshift, and the parametric fit to the mass function for star-forming galaxies given by Muzzin et al.\ (2013b).

For each redshift bin we build a stack of radio flux densities by taking the value of the pixel in the VLA-COSMOS image that corresponds to each position. In addition to the single peak pixel stack, we also build a 2D image stack. This 2D stack is used to assess the integrated-to-peak flux density conversion ratio for each set of objects. For point sources the integrated-to-peak flux density ratio should be unity, however some of the sources in our stacks may be extended (especially at low-$z$), and bandwidth smearing at the edge of the VLA pointings has the effect of making sources marginally extended. To determine the correct integrated-to-peak flux density ratio for each of our stacks we produce $501\times501$ pixel stacks for each of our redshift bins. The median image is produced across these stacks, and then de-convolved with the dirty beam. The peak flux density is then compared to the integrated value across the median image. The integrated-to-peak flux density ratios ($S_{\rm int}/S_{\rm peak}$) for each redshift bin assessed in this way are given in Table \ref{tab:stackbins}. It is interesting to compare our values to those of Karim et al.\ (2011; K11), who performed a similar stacking analysis of the VLA-COSMOS data. While they use a slightly different map (the COSMOS internal Deep map, rather than the public Large Project map), and redshift binning, the trend of large $S_{\rm int}/S_{\rm peak}$ at low-$z$ decreasing to a typical value of $\sim1.7$--$2$ at $z>0.5$ is also seen.

After correcting the pixel stack values by $S_{\rm int}/S_{\rm peak}$, we fit the model specified by equations \ref{Eqn:pdm} and \ref{Eqn:pfz} using Monte-Carlo Markov Chain (MCMC) methods. Specifically, we use an implementation of the affine-invariant sampling proposed by Goodman \& Weare (2010). A detailed discussion of this MCMC approach, and its application to astronomy, can be found in Foreman-Mackey et al.\ (2013), we briefly summarise the main points here. The affine invariant sampling method makes use of a set of test positions, or ``walkers", on the posterior probability. In each iteration the proposed ``step'' for a walker is generated by defining a vector in parameter space between it and another randomly chosen ``walker" and moving along this vector by a random fraction between $1/a$ and $a$, where $a$ is a tuning parameter typically set to two. Proposed steps are then accepted with a probability such that MCMC chain is ``balanced" (i.e. it is equally likely to step from $x\rightarrow x^{\prime}$ as its inverse $x^{\prime}\rightarrow x$). Here we use our own {\sc IDL} implementation of the affine-invariant ensemble sampler; for each stack we utilise 200 walkers and the tuning parameter $a=2$. The correlation length (i.e. the number of steps required for the chain to retain no knowledge of its starting point) for each parameter is measured using the method of Goodman \& Weare  (2010), and the MCMC chain run until the number of steps exceeds ten times the maximum correlation length amongst the all of the free parameters.

The model described in \S\ref{sec:stackmethod}  and \S\ref{sec:data} contains five free parameters: The Gaussian noise in the radio image, $\sigma_{N}$; the LF normalisation, $\Phi^{\ast}$; the characteristic luminosity, $L^{\ast}$; the faint end slope, $\alpha$; and the Gaussian term in the LF, $\sigma_{\rm LF}$. For the parameters $\sigma_{N}$ and $\sigma_{\rm LF}$ we assume a Gaussian prior, while for the others we assume a ``top-hat'' prior with an arbitrarily chosen range. A summary of the model parameters and their priors is given in Table \ref{tab:modelpriors}. For the image noise, $\sigma_{N}$ we set the prior to the value obtained from the standard deviation of all the pixels in radio image within the ULTRAVISTA coverage, giving $\sigma_{N}=13.6\pm1$. Meanwhile, the prior for the LF parameter, $\sigma_{\rm LF}$, is set to the best estimate and uncertainty from fitting Eqn.~\ref{Eqn:msf} to the measurements of the local radio LF for star-forming galaxies from Condon et al.\ (2002) and Mauch et al.\ (2007). 

\begin{table}
\caption{Details of parameters used to fit pixel stacks}
\label{tab:modelpriors}
\begin{tabular}{l|llll}
Parameter & Min & Max & $\mu$ & $\sigma$\\
\hline
$\sigma_{\rm N}^1$ & 0 & 50 & 13.6 & 1\\
$\Phi^{\ast}$ & -10 & 12 & -- & --\\
$L^{\ast}$ & 18 & 25 & -- & -- \\
$\alpha$ & 0 & 1 & -- & --\\
$\sigma_{\rm LF}^1$ & 0 & 2 & 0.6 & 0.1\\
\hline
\multicolumn{3}{l}{$^1$Gaussian prior}\\
\end{tabular}
\end{table}

\section{Results}\label{sec:sresults}
\begin{figure*}

\includegraphics[scale=0.6]{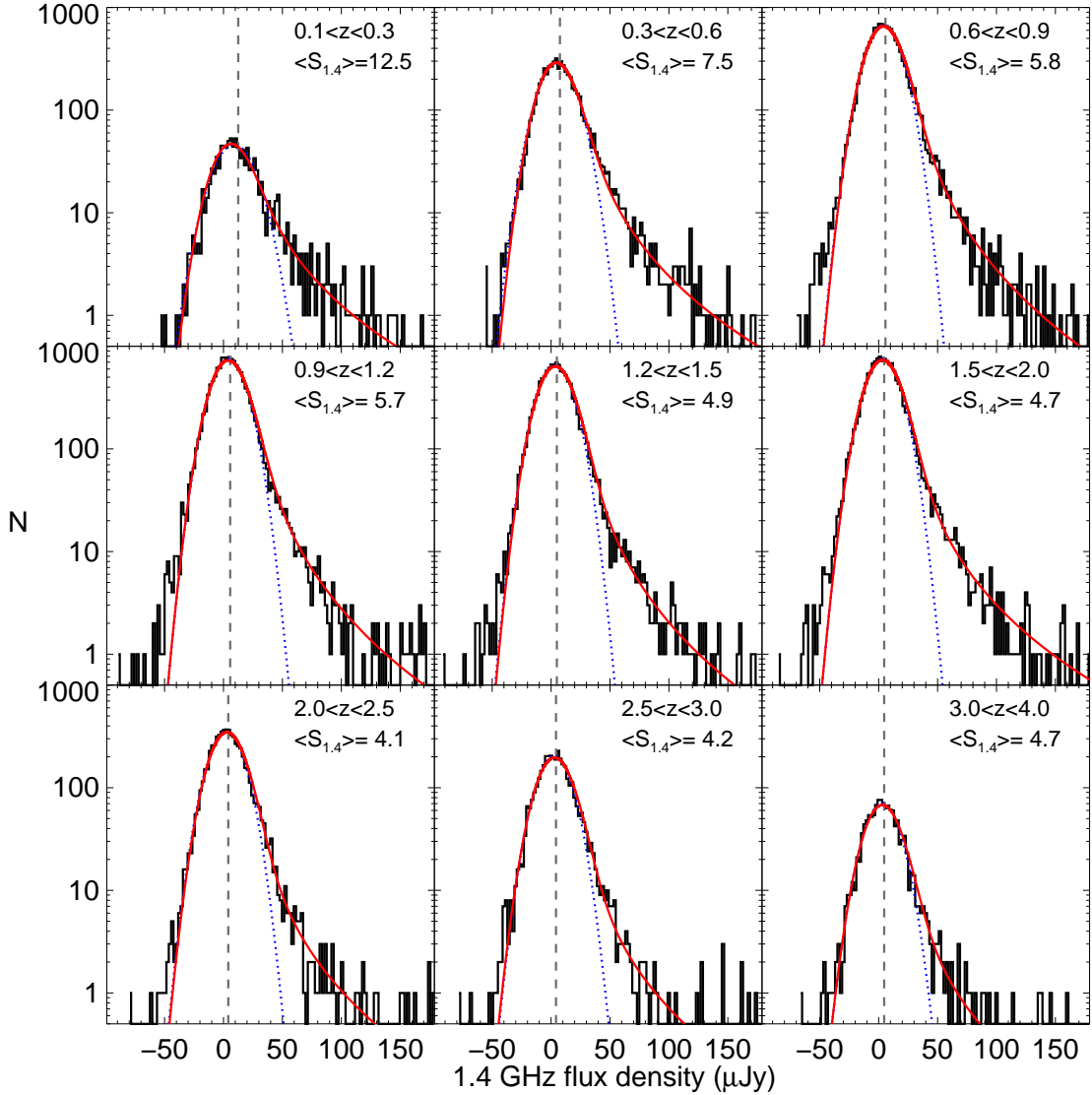}
\caption{Pixel flux density distribution compared to both the median-fit model (red solid line) and a simple Gaussian fit (blue dotted line). The median of the flux density distribution is shown as a grey dashed line and quoted in the top right corner. In each case it can be seen that there is both a non-zero median of the flux density distribution, and an excess of pixels in the stack with positive values over the simple Gaussian fit, validating the basic assumptions of our method.}
\label{fig:stacks}
\end{figure*}

The median stack-fit estimates and 68 per cent confidence intervals for the model parameters are given in Table \ref{tab:stackbins}. Figure \ref{fig:stacks} shows the distribution of flux density in stack pixels for the nine redshift bins compared to both the median stack-fit and a simple Gaussian fit to the pixel stack. Also shown is the median of the flux density distribution, which is always found to take a positive value. In each redshift bin there is an excess of pixels with positive values over the Gaussian model, validating our basic model assumption.

\begin{table*}
\caption{Details of the stacking bins, and median fit values to the stacks achieved via MCMC. In each case the quoted errors represent the 68 per cent confidence interval.}
\label{tab:stackbins}
\begin{tabular}{llllllllll}
\hline
$z$ & $N$ & $\langle S_{1.4}\rangle$ & $S_{\rm int}/S_{\rm peak}$ & $\sigma_{N}$ & $\Phi^{\ast}$ & $L^{\ast}$ & $\alpha$ & $\sigma_{\rm LF}$\\
\hline
& & $\mu$Jy & & $\mu$Jy & Mpc$^{-3}$ & W\,Hz$^{-1}$ &  & W\,Hz$^{-1}$\\
\hline  
$0.1$--$0.3$ & 1083 & 12.5 & 3.6 & $13.8^{+0.5}_{-0.5}$      & $-2.72^{+0.10}_{-0.19}$ & $20.4^{+0.4}_{-0.3}$ & $0.36^{+0.21}_{-0.25}$ & $0.63^{+0.05}_{-0.06}$\\
$0.3$--$0.6$ & 5585 &  7.5 & 2.65 & $13.0^{+0.2}_{-0.2}$   & $-2.86^{+0.14}_{-0.13}$ & $20.7^{+0.3}_{-0.2}$ & $0.19^{+0.20}_{-0.14}$ & $0.63^{+0.02}_{-0.03}$\\
$0.6$--$0.9$ & 12095 & 5.8 &  1.88 & $13.0^{+0.1}_{-0.1}$ & $-2.79^{+0.08}_{-0.06}$ & $21.2^{+0.1}_{-0.1}$ & $0.08^{+0.12}_{-0.06}$ & $0.58^{+0.02}_{-0.02}$\\
$0.9$--$1.2$ & 13319 & 5.7 & 1.70 & $13.0^{+0.1}_{-0.1}$  & $-2.92^{+0.06}_{-0.05}$ & $21.4^{+0.1}_{-0.1}$ & $0.06^{+0.08}_{-0.04}$ & $0.58^{+0.02}_{-0.02}$\\
$1.2$--$1.5$ & 11517 & 4.9 &1.80 & $13.0^{+0.1}_{-0.1}$  & $-3.05^{+0.10}_{-0.08}$ & $21.6^{+0.2}_{-0.1}$ &  $0.12^{+0.15}_{-0.09}$ & $0.60^{+0.02}_{-0.02}$\\
$1.5$--$2$ & 13376 & 4.7 & 1.76 & $13.1^{+0.1}_{-0.1}$    & $-3.26^{+0.14}_{-0.12}$ & $21.7^{+0.3}_{-0.2}$ &  $0.16^{+0.18}_{-0.12}$ & $0.67^{+0.02}_{-0.02}$\\
$2$--$2.5$ & 6255 & 4.1 & 1.90 & $13.2^{+0.2}_{-0.2}$       & $-3.41^{+0.10}_{-0.18}$ & $22.1^{+0.4}_{-0.3}$ &  $0.32^{+0.23}_{-0.23}$ & $0.65^{+0.03}_{-0.04}$\\
$2.5$--$3$ & 3606 & 4.2 & 1.98 & $13.4^{+0.2}_{-0.2}$       & $-3.60^{+0.08}_{-0.22}$ & $22.3^{+0.5}_{-0.4}$ &  $0.38^{+0.25}_{-0.27}$ & $0.67^{+0.04}_{-0.05}$\\
$3$--$4$ & 1257 & 4.7 & 1.92 & $13.3^{+0.4}_{-0.4}$          &  $-4.29^{+0.08}_{-0.22}$ & $22.6^{+0.5}_{-0.5}$ & $0.41^{+0.26}_{-0.29}$ & $0.67^{+0.06}_{-0.07}$\\
\hline
\end{tabular}
\end{table*}

While the degeneracies between the model parameters make direct interpretation of the median stack-fit parameters problematic, there are some noteworthy trends. Reassuringly, the estimates of the noise in the map are consistent across the nine redshift bins. Some real variation in this quantity is to be expected due to the variation in $S_{\rm int}/S_{\rm peak}$ from both extended sources and bandwidth smearing. Apart from the lowest redshift bin (where we would expect many sources to be resolved) the variation in $\sigma_{N}$ is within the measurement errors. 

The relationship between $\Phi^{\ast}(z)$ and $L^{\ast}(z)$ is interesting. At $z\ls1$, $\Phi^{\ast}(z)$ remains roughly constant, while $L^{\ast}(z)$ increases steadily, i.e. pure luminosity evolution. This picture is consistent with simple models for the evolution of star forming galaxies in the radio from previous work (e.g. Haarsma et al.\ 2000; Smolcic et al.\ 2009). At $z\gs1$, $\Phi^{\ast}(z)$ decreases rapidly while $L^{\ast}(z)$ continues to increase strongly to our limit of $z\sim4$. While radio LF measurements for star-forming galaxies do not currently exist above $z=1$, this evolution in the LF parameters is in agreement with the known trend for star forming galaxies from other tracers at high redshift (e.g.\ Gruppioni et al.\  2013; Sobral et al.\ 2012).

\section{Discussion}\label{sec:discussion}
\subsection{Luminosity functions and the cosmic star formation rate density.}\label{sec:sfrd}
\begin{figure*}
\includegraphics[scale=0.6]{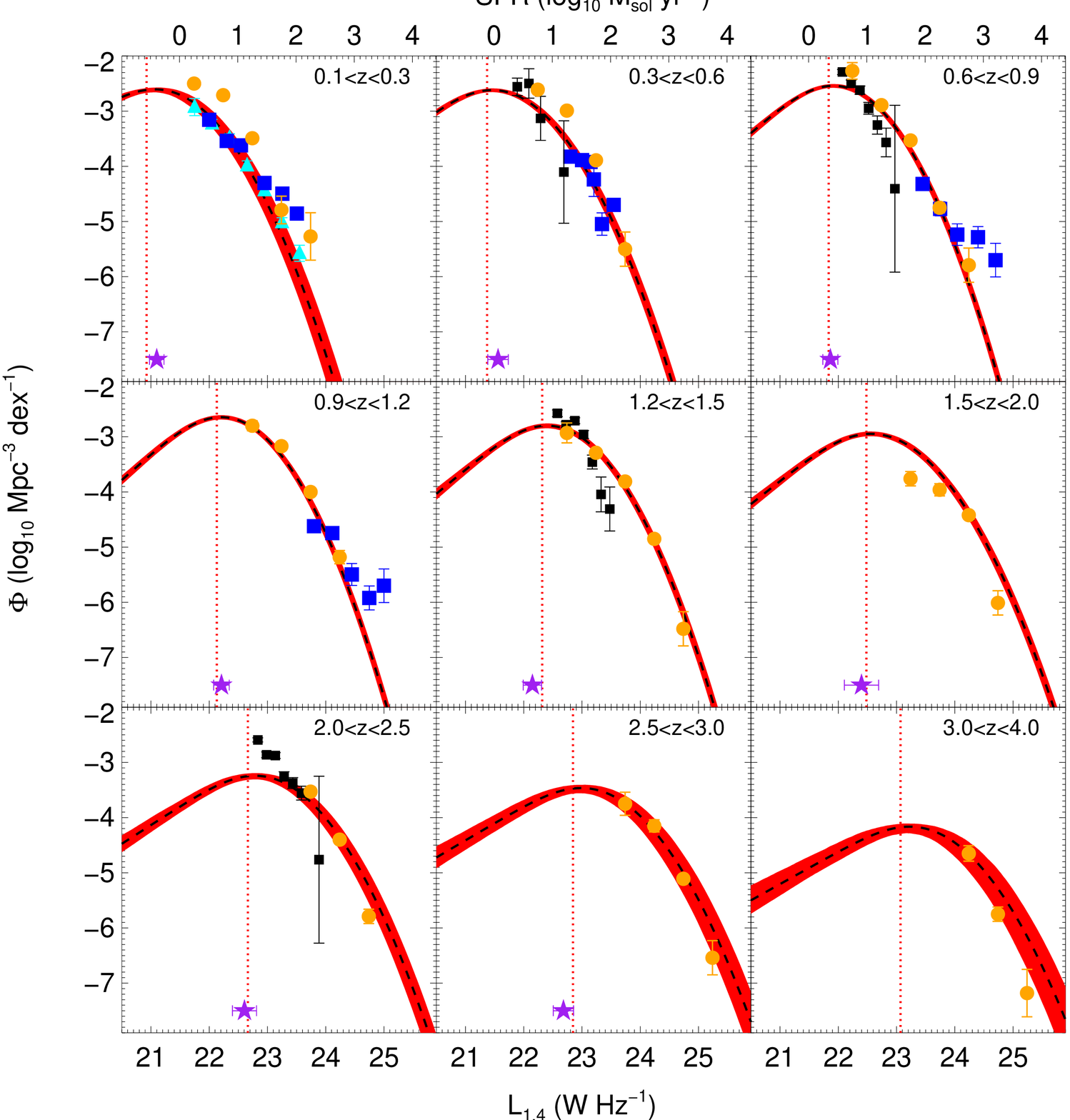}
\caption{Luminosity functions from stack-fitting to the radio pixel intensity stacks. The solid dashed line is the LF from the median fit parameters given in Table \ref{tab:stackbins} while the red shaded region is the 68 percent interval in $\Phi$ at each $L_{1.4}$.  Literature values of the radio LF from other works are also shown as: cyan triangles (NVSS-SDSS; Best et al.\ 2005) and blue squares (VLA-COSMOS; Smolcic et al.\ 2009). Far-IR and mass-limited H$\alpha$ estimates, converted to radio luminosity assuming star-formation as the common origin, are shown as orange circles ({\it Herschel} Far-IR; Gruppioni et al.\ 2013) and black squares (H$\alpha$; Sobral et al.\ 2013). The purple star is the median $L_{1.4}$ found in equivalent mass and redshift bins by K11, while the dotted line represents the median of our parametric fit.}
\label{fig:lfplots}
\end{figure*}

The stack-fit mass-limited radio luminosity functions for star forming galaxies are shown in Fig.~\ref{fig:lfplots}. The 68 per cent confidence interval on our LF estimates (and subsequently SFRd estimates) are constructed by generating Monte-Carlo realisations of our best-fit parameters, using the covariance matrix output from the MCMC chains. In each realisation the normalisation of the LF is forced to equal the completeness corrected number of sources in the field, with an additional noise term taking into account both Poisson noise and cosmic variance. The effect of cosmic variance is estimated using the results of Moster et al.\ (2011), and found to be 6 per cent for the lowest redshift bin, increasing to 18 per cent for the highest redshift bin.

Directly comparing our LF stack-fits to literature values is complicated by the fact that our sample is stellar mass limited, rather than radio luminosity limited. Nonetheless, we compare our results to total LF's in the radio  from Best et al.\ (2005) and Smolcic et al.\ (2009), the latter of which makes use of the same VLA COSMOS imaging as this work.  The two sets of radio luminosity functions show good agreement with our estimates, although the two highest redshift bins probed by Smolcic et al. (2009) suggest an upturn at $L_{1.4}\gs10^{24.5}\,$W\,Hz$^{-1}$ that is not possible for our parametric fits to replicate. 

It is also interesting to compare our results to the median radio luminosities found via traditional stacking by K11. For each redshift bin in Fig.~\ref{fig:lfplots} we show the median radio luminosity for $M_{\ast}>10^{9.5}\,$M$_{\odot}$ from K11 as a purple star, while the median luminosity of our parametric fit is shown as a red dashed line. It can be seen that the K11 values agree well with the median luminosity of our LFs, both validating our approach and also demonstrating the power of our parametric stack-fitting over traditional stacking techniques.

Finally, we compare our results to the total far-IR LF (Gruppioni et al.\ 2013) and the mass-limited H$\alpha$ luminosity function (Sobral et al.\ 2013). Assuming in all cases that star-formation dominates the luminosity, we can convert the Far-IR and H$\alpha$ measurements to 1.4\,GHz luminosity by combining the SFR calibrations of Kennicutt et al.\ (1998) and Bell et al.\ (2003); $L_{1.4}=L_{IR}+11.49$ and $L_{1.4}=L_{H\alpha}-19.87$, assuming a Chabrier (2003) IMF. Excellent agreement is found between our estimate and the far-IR measurements from Gruppioni et al.\ (2013) at $z>0.6$, while lower redshift bins appear to show an `excess' of far-IR sources over the radio estimates. This result is not unique to our work, and has been noted by previous authors when comparing radio and far-IR LFs for star-forming galaxies (e.g. Bell et al.\ 2003; Smolcic et al.\ 2009).


The comparison with the H$\alpha$ estimates is especially interesting as these measurements have been made with the same mass-limit as our sample. For the $0.3<z<0.6$ bin the agreement between the H$\alpha$ and radio mass-limited LF is very good. At higher redshifts the overall normalisation of the H$\alpha$ and radio LFs agree well, but the H$\alpha$ shows a steeper cut-off towards high luminosities (i.e. lower $L^{\ast}$). This could point to insufficient dust corrections to the H$\alpha$ fluxes for galaxies with the highest star formation rates (e.g. Roseboom et al.\ 2012). Alternatively, our radio measurements could be polluted with emission not related to star formation (e.g. AGN). However, the agreement seen between the radio and far-IR at high luminosities means that the far-IR estimates also need to be influenced in the same way. The physical origin of far-IR emission in ULIRGs (and by proxy the radio) is an area of some debate, although most studies agree that it is very difficult to power luminous IR emission at long wavelengths (i.e. $\gs100\,\mu$m) with a process other than star formation (e.g. Lutz et al.\ 2008; Hatziminaoglou et al.\ 2010).


\begin{figure}
\includegraphics[scale=0.45]{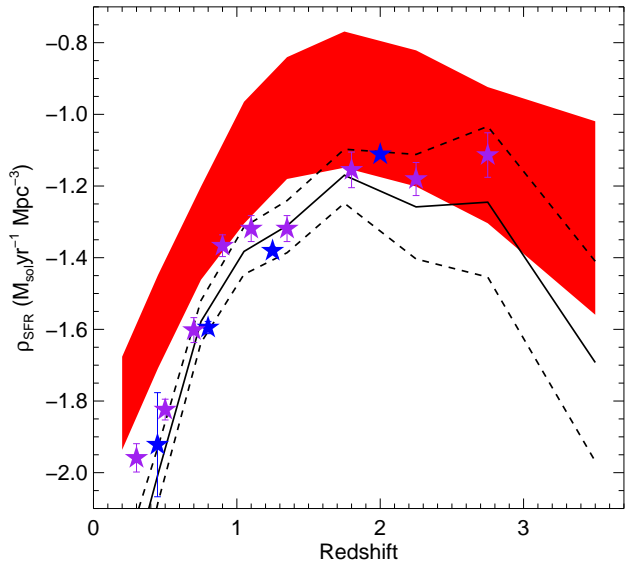}
\caption{Star formation rate density (SFRd) for galaxies with $M_{\ast}>10^{9.5}\,$M$_{\odot}$ as a function of redshift implied by our luminosity functions (solid line). The dashed lines represent the 1$\sigma$ uncertainty on our parametric fit. For comparison is shown the best estimate of the total SFRd(z) from the compilation of Behroozi et al.\ (2013; red shaded), as well as mass-limited estimates from both the UV (Santini et al. 2009; blue circles) and stacked radio (K11; purple stars). Both our parametric fits, and the literature results, imply that $M_{\ast}\gs10^{9.5}\,$M$_{\odot}$ galaxies contribute $\gs50$ percent of the total SFRd up to $z<4$}
\label{fig:sfrdz}
\end{figure}

Armed with our LFs for $M_{\ast}>10^{9.5}\,$M$_{\odot}$ galaxies out to $z=4$, and assuming the radio luminosity is generated by star-formation, we can ask the question; what contribution do these galaxies make to the cosmic SFR density (SFRd)? In Fig.~\ref{fig:sfrdz} we show the cosmic SFRd estimated by integrating the LFs  from Fig.~\ref{fig:lfplots} and assuming the SFR calibration of Bell et al.\ (2003). Also shown in Fig.~\ref{fig:sfrdz} is the recent compilation of total SFRd estimates by Behroozi et al.\ (2013; B13), as well as a mass-limited ($M_{\ast}>10^{9.8}\,$M$_{\odot}$) estimate constructed from the results of Santini et al. (2009). While the error in our estimate, and the scatter in the B13 literature values, are significant, it is clear that galaxies with $M_{\ast}>10^{9.5}\,$M$_{\odot}$ make up $\gs50$ percent of the SFRd up to $z=4$.  This is in good agreement with both the K11 and Santini et al. (2009) results up to $z\sim3$. 

\subsection{Future prospects for parametric stack-fitting}\label{sec:future}
From the results above it is clear that stack-fitting has substantial benefits if source catalogues at one wavelength are significantly deeper compared to overlapping data and a flexible parametric description of the data can be provided. As we have shown, this technique is well-suited to interferometric radio data, as the disparity in angular resolution compared to optical/near-IR imaging means that source confusion is not an issue.

Several large surveys of the extragalactic sky at radio wavelengths are either underway (e.g. with LOFAR; van Haarlem et al.\ 2013) or will commence in the near future (e.g. MIGHTEE, Jarvis et al.\ 2011; EMU, Norris et al.\ 2012). These surveys will provide all-sky observations at comparable depth to the COSMOS data, as well as significantly deeper regions of comparable area. However, none of these surveys will provide radio imaging that is effectively deeper than ancillary optical/near-IR datasets. Thus, techniques such as the one presented here may be invaluable in terms of interpreting these surveys. 

At other wavelengths, large areas of the sky have already been imaged at IR/submm wavelengths by {\it Herschel}, {\it WISE} and SCUBA-2. Again, overlapping optical/near-IR datasets are significantly deeper and so the stack-fitting approach may offer significant benefits, with the caveat that the mismatch in the beam size (roughly a factor of 10 between ground-based near-IR and submm) means that confusion noise must be taken into account, something which is not currently accommodated by our method. 

While we leave a detailed treatment of implementing our method in the presence of confusion noise to future work, we briefly consider here how confusion noise could be accounted for. The simplest solution to the problem of source confusion is to assume that the confusion introduces some noise term, similar to the instrumental noise in Eqn.~\ref{Eqn:pdm}, which is a function of both the instrument beam and the intrinsic flux density distribution, 

\begin{equation}
P^{\rm conf}(f)= P(f) + \int_{s=0}^{\infty}\int_{r=0}^{\infty}B(r)sP(s)\d{r}\,\d{s},
\label{Eqn:confpf}
\end{equation}

\noindent where $B(r)$ is the 1-dimensional beam profile, and $P(s)$ is the number density of sources with intrinsic flux density $s$. An implicit assumption of Eqn.~\ref{Eqn:confpf} is that the sources are randomly distributed on the sky. In the presence of clustering, if the sources in the stack can be described by a simple model (e.g. a power-law, or halo occupation model), this can be added to the integral such that the probability of source occurring within the beam is also a function of radius from the source, i.e. $P(s)$ becomes $P(s,r)$.

Finally, it is of interest to understand how our method scales with the number of sources ($N_{\rm s}$) in the stacks. For traditional stacking the signal-to-noise in the stacked signal increases as $\sqrt{N_{\rm s}}$. However, it is not obvious how the precision in our model LF parameters should vary as a function of $N_{\rm s}$. To test this we take the redshift bin with the most sources, $1.5<z<2$, from our analysis and sparsely sample it to simulate the impact of decreasing the number of sources. Details of this process are given in Table~\ref{tab:testn}.

\begin{table}
\caption{Effect of varying $N_{\rm s}$ on the precision of the LF parameters, $L^{\ast}$ and $\alpha$, and the area 95 per cent confidence ellipse between these parameters, $A_{95}$.}
\label{tab:testn}
\begin{tabular}{llll}
$N_{\rm s}$ & $\sigma_{L^{\ast}}$ & $\sigma_{\alpha}$ & $A_{95}$ \\
\hline
13376 & 0.21 & 0.14 & 0.19\\
9515 & 0.25 & 0.16 & 0.26\\
4577 & 0.28 & 0.18 & 0.36\\
1828 & 0.31 & 0.18& 0.56\\
935 & 0.39 & 0.21& 0.90\\
468 & 0.36 & 0.20 & 1.09\\
182 & 0.72 & 0.27 & 1.55\\
\hline
\end{tabular}

\end{table}

One difficulty in the analysis of this simulation is that the parameters in the model are degenerate, so the individual error estimates do not fairly represent the true uncertainty. To mitigate this we consider the area of the 1$\sigma$ error ellipse defined by the covariance matrix for each model fit. As $\sigma_N$ and $\sigma_{\rm LF}$ are constrained by Gaussian priors (and only weakly degenerate with $L^{\ast}$) and $\Phi^{\ast}$ is determined from the absolute normalisation, we need only consider the ellipse formed by the uncertainty in $L^{\ast}$ and $\alpha$. Thus we also list the area of the ellipse which contains 95 percent of the uncertainty in $L^{\ast}$ and $\alpha$, $A_{95}$, in Table~\ref{tab:testn}.

For traditional stacking techniques the precision in the stacked estimate increases $\propto N^{1/2}$. To allow quantitative comparison with these techniques we fit the data from Table~\ref{tab:testn} with a model assuming the precision, $\Delta$, improves with the number of sources to some fractional exponent i.e. $\Delta\propto N_{\rm s}^{-1/\tau}$. This gives $\tau=4.2\pm0.7$ for $\sigma_{L^{*}}(N_{\rm s})$,  $\tau=8\pm2$ for $\sigma_{\alpha}(N_{\rm s})$ and $\tau=2\pm0.1$ for the 95 percent confidence ellipse area ($A_{95}$). It is clear that the large degeneracies between the parameters significantly inflate the exponents for the marginalised parameter uncertainties. However,  the area of the error ellipse between $L^{\ast}$ and $\alpha$ is seen to scale as $N_{\rm s}^{-1/2}$, i.e. as the square-root of the number of sources in the stack, similar to simple stacking methods.

\section{Conclusions}\label{sec:conc}
In the last decade there has been an increasing focus on multi-wavelength approaches to outstanding problems in galaxy formation and evolution. While multi-wavelength tracers of galaxy parameters (e.g. star formation rate) are key in reducing systematic uncertainties, the disparity in observing capabilities across wavelength regimes, in terms of both sensitivity and angular resolution, severely limits the ability to build cross-matched samples of individual galaxies. In this work we have presented an alternative method designed to maximise the information gleaned from multi-wavelength datasets by using a statistical approach to constrain a parametric description of the luminosity function. Taking a mass-limited sample in the COSMOS field, and stack-fitting pixel intensity distributions from the VLA-COSMOS imaging, we achieve the following results:
\begin{itemize}
\item We constrain the parameters of the mass-limited radio LF for star-forming galaxies out to $z\sim4$, with $\sim0.2$\,dex error in $L^{\ast}$.
\item Our stack-fit LFs show good agreement with comparable literature estimates. Compared to direct estimates using the same radio data (VLA-COSMOS) we can constrain the LF 1--2\,dex below what is possible with individually detected sources at $z>0.3$. 
\item Converting our radio LFs to star formation rate density (SFRd), we find that $M_{\ast}>10^{9.5}\,$M$_{\odot}$ galaxies make up $\gs50$ percent of the total SFRd in the interval $0.1<z<4$, in good agreement with other estimates from the optical and mid-IR.
\end{itemize}

Given the wealth of existing multi-wavelength data at far-IR/submm wavelengths, and the promise of large-scale radio continuum surveys in the near future leading to the Square Kilometer Array (SKA), it is clear that statistical methods such as the one presented here will be invaluable in determining the cosmic history of star forming galaxies. As a final result we show that the precision of our method scales like $\sqrt{N_{\rm s}}$, where $N_{\rm s}$ is the number of sources used in the pixel stack. Thus it is clear that even with large area, shallow, radio surveys such as those proposed for the SKA pathfinders ASKAP and MeerKAT (EMU, Norris et al.\ 2012; MIGHTEE, Jarvis et al.\ 2011) it will be possible to obtain interesting results about the very high redshift Universe well below the nominal detection limits for individual sources.

\section*{Acknowledgements}
We thank the anonymous referee for their constructive comments which enhanced this work. We would also like to thank J.~Peacock, A.~Karim and J.~Dunlop for useful conversations which helped this work, E.~Schrinnerer for providing the VLA-COSMOS dirty beam map, A.~Muzzin \& D.~Marchesini for both making the ULTRAVISTA catalogue public and providing the catalogue completeness curves, and D.~Sobral for providing the mass-limited H$\alpha$ LF estimates.

\label{lastpage}

\end{document}